# Bose Einstein correlations of neutral pion pairs at LEP

M. Boutemeur[1] and G. Giacomelli[2]

Talk Given by G. Giacomelli at DPF 2003, Philadelphia USA, 5$^{th}$ – 8$^{th}$ april 2003

[1]Ludwig-Maximilians-Universität München, Am Coulombwall 1,
D-85748 Garching, Germany; Madjid.Boutemeur@cern.ch

[2]Dipartimento di Fisica of the University of Bologna and INFN Sezione di Bologna,
I-40127 Bologna, Italy; Giacomelli@bo.infn.it

**Abstract.** With the OPAL detector at LEP we measured at energies around the $Z^0$ peak the Bose-Einstein Correlations (BECs) of neutral pion pairs. We compare the results of this measurement with former results obtained at LEP for hadrons including those obtained from Fermi-Dirac Correlations (FDCs).

**Introduction:** Bose-Einstein Correlations (BEC) are a quantum mechanic phenomenon, which arises from the ambiguity of path between sources and detectors and from the requirement to symmetrize the wave function of two or more identical bosons. They manifest as an enhanced probability for identical bosons to be emitted with small relative four momentum Q, compared with non identical bosons under similar kinematic conditions [1, 2, 3]. From the measured effect it is possible to determine the space time dimensions of the boson source and the chaoticity parameter. Many experimental studies have been performed for identical charged pions and kaons, and few on neutral pion pairs. In this note we shall concentrate on a new measurement of BECs of $\pi^0\pi^0$ in multihadronic events at LEP1 energies with the OPAL detector [4]. We shall then compare these results with those of previous measurements of charged and neutral hadron pairs, of multiple hadron BECs, and of Fermi-Dirac Correlations (FDC) for charged and neutral fermions. Contrary to BEC, FDC effects manifest as a lack of events at small relative four momenta.

**Data selection:** The OPAL detector is described in [5]. For the study of $\pi^0\pi^0$ BECs we used only the central tracking detector and the barrel electromagnetic calorimeter [4]. A sample of 3.1 million $e^+e^-$ multihadron annihilations at energies around the $Z^0$ peak were used [5]. Further selections were made: 1) Hadronic events were required to have more than 7 tracks and to form two well defined back-to-back jets. 2) Photons with energy larger than 0.1 GeV are paired to form $\pi^0$ candidates where each photon pair is required to have a probability to yield a $\pi^0$ larger than 0.6. 3) In order to remove $\pi^0$s produced in the detector material, the momentum of each $\pi^0$ candidate is required to be larger than 1 GeV/c. 4) Only events with one possible $\pi^0$ pair from four distinct photons were retained for the analysis.

Fig 1 shows the two photon invariant mass distribution. In the selected window (100-170 MeV ) the $\pi^0$ purity is 79%. The purity of the $\pi^0\pi^0$ sample is 60%.

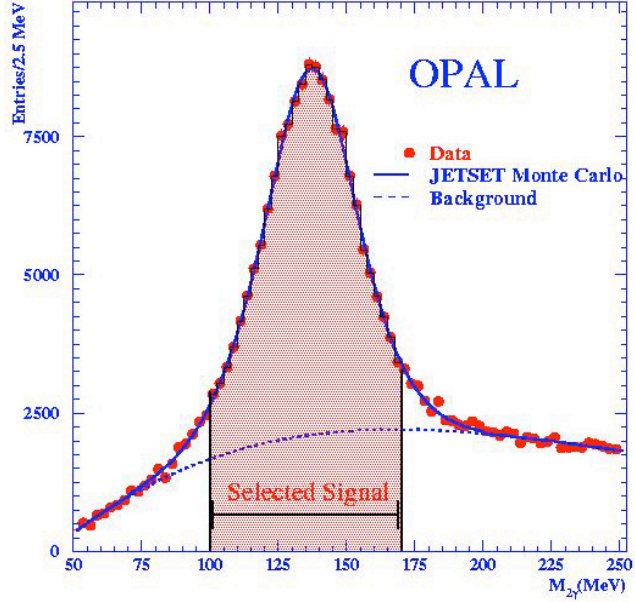

Fig. 1: Distribution of two-photon invariant mass, $M_{2\gamma}$, for selected events. The smooth curves are the total Monte Carlo expectation (solid line) and the background expectation (dashed line). The shaded region (100 – 170 MeV) is the selected window for the $\pi^0$ signal.

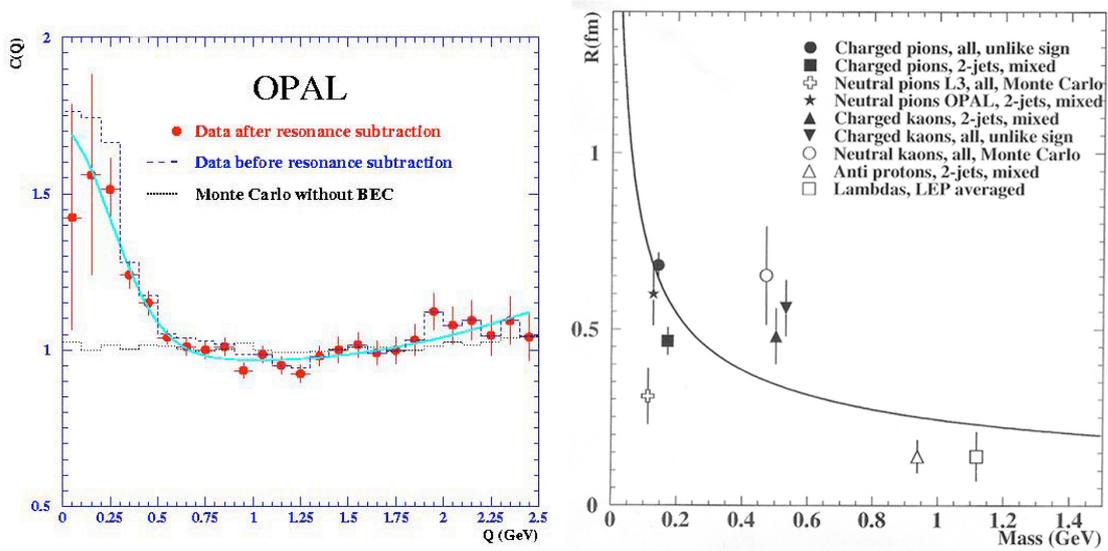

Figure 2: The BEC distribution $C(Q)$ for the $\pi^0\pi^0$ data. The smooth curve is the fitted correlation function, the dotted histogram is the distribution from JETSET Monte Carlo events generated without BEC. The dashed histogram is the measured correlation function before subtraction of known hadron decays.

Figure 3: Radius of the emitting region for BEC for two identical bosons and for FDC for two identical baryons produced in $e^+e^-$ collisions at LEP, plotted vs hadron mass.

**Analysis method:** The BEC function is defined as the ratio $C(Q) = \rho(Q)/\rho_0(Q)$, were $Q$ is a Lorenz-invariant variable expressed in terms of the two $\pi^0$ four momenta $p_1$ and $p_2$ as $Q^2 = -(p_1 - p_2)^2$, $\rho(Q) = (1/N)\, dN/dQ$ is the measured $Q$ distribution of the two $\pi^0$s and $\rho_0(Q)$ is a reference distribution which should contain all the correlations included in $\rho(Q)$ except BECs. For the determination of $\rho_0(Q)$, we considered two methods [4]: the event mixing reference sample, where mixed $\pi^0$ pairs are formed from $\pi^0$s belonging to different events in the data, and the Monte Carlo (MC) reference sample where the $\rho_0$ distribution is obtained from a Monte Carlo simulation without BEC. Here, the mixing technique is used as the main analysis method and the Monte Carlo reference technique is applied only for comparison.

**Results:** The correlation distribution $C(Q)$ is parametrised using the Fourier transform of the expression for a static sphere of emitters with a Gaussian density:
$$C(Q) = N(1 + \lambda \exp(-R^2 Q^2))(1 + \delta Q + \varepsilon Q^2). \quad (1)$$
$\lambda$ is the chaoticity parameter, $R$ is the radius of the source, and $N$ a normalization factor. The empirical term, $(1 + \delta Q + \varepsilon Q^2)$, accounts for the behaviour of the correlation function at high $Q$ due to any remaining long-range correlations. The measured $C(Q)$ distribution for the data is shown in Fig. 2 as the black points with corresponding statistical errors; the smooth curve is the fitted correlation function in the $Q$ range $0 < Q < 2.5$ GeV. A clear BEC enhancement is observed in the low $Q$ region of the distribution. The least square fitted parameters to Eq. 1 are:
$\lambda = 0.55 \pm 0.10(stat.) \pm 0.10(syst.)$,
$R = (0.59 \pm 0.08(stat.) \pm 0.05(syst.))$ fm,
$N = 1.10 \pm 0.08(stat.)$,
$\delta = (-0.14 \pm 0.05(stat.))$ GeV$^{-1}$,
$\varepsilon = (0.07 \pm 0.03(stat.))$ GeV$^{-2}$,
where the $\chi^2/d.o.f$ of the fit is 14.7/19. Potential sources of systematic errors are investigated. In each case the effect on the measured parameters and their deviations with respect to the standard analysis are estimated. The final systematic errors, quoted above for $R$ and $\lambda$, are obtained from quadratically adding the deviations from the central value. For comparison, the second method which uses the MC reference sample yields the following results: $\lambda = 0.50 \pm 0.10(stat.)$, $R = (0.46 \pm 0.08(stat.)$ fm with $\chi^2/d.o.f$ of 15.1/19.

**Discussion:** Since in multihadron events more than 90% of the measured tracks are charged pions, the study of BEC for like-sign charged pion pairs was usually performed without proper particle identification and without purity correction. This choice introduces an error in the chaoticity parameter and in the radius, $R$, of the emitting region. However some analyses were performed with properly identified pions. That required some effective cuts on the fraction of the global solid angle acceptance. For $\pi^0\pi^0$, $K^\pm K^\pm$ and $K^0 K^0$ correlations, the identification of the particles was necessary [4, 6]. The largest difference among the results from different experiments lies essentially in the choice of the reference sample. Thus while the statistical errors on $R$ can be small, the systematic uncertainty is large.

Fig. 3 shows the radius $R$ for BEC ($\pi^0\pi^0$, $\pi^\pm\pi^\pm$, $K^\pm K^\pm$, $K^0 K^0$) and for FDC (for $\bar{p}\bar{p}$, $\Lambda^0\Lambda^0$). The graph may be illustrative of the present situation: (i) There could be a decrease of R with increasing mass of the particles considered; this is the preferred interpretation in [7]. But (ii) there could be also a small difference between $\pi^0\pi^0$ and $\pi^\pm\pi^\pm$, with $R(\pi^\pm\pi^\pm) > R(\pi^0\pi^0)$. (iii) Probably more favoured by Fig. 3, the radius for pion and kaon pairs could be similar $R(\pi^0\pi^0) \cong R(\pi^\pm\pi^\pm) \cong R(K^\pm K^\pm) \cong R(K^0 K^0) \cong 0.64$ fm [6], while that from the FDC is definitely smaller with $R(\bar{p}\bar{p}) \cong R(\Lambda^0\Lambda^0) \cong 0.14$ fm.

In ref. [8] it was shown that there is an increase of about 10% of the emitting region radius when the multiplicity increases from 10 to 40 charged hadrons in the final state. This could be related to the number of hadron sources, i.e. the number of jets. The study of BEC in multidimensional space has indicated that the emitting source is ellipsoidal, with the transverse radius $R_t$ smaller than the longitudinal radius $R_l$, more precisely $R_t \cong 0.8\, R_l$ [9].

In ref [10] it was shown that there are genuine $3\pi$ BEC, that is after removing the effect of $2\pi$ correlations on the $3\pi$ sample. The present situation is consistent with the relation $R_{\pi^\pm\pi^\pm\pi^\pm} \cong R_{\pi^\pm\pi^\pm} / \sqrt{2}$.

For completeness, BEC should really be studied in four dimensions, the fourth dimension being time. There are indications that the time involved in the emission process is of the order of $10^{-24}$ s, comparable to the hadronization time.

**Conclusions:** At the present stage, BEC and FDC studies cannot be considered as precision measurements. However, with the already available results, one could hope for a comprehensive model which could at least help in implementing properly the effect in the Monte Carlo simulations.

**Aknowledgements**. We would like thank the members of the OPAL Collaboration for their cooperation, in particular C. Ciocca and M. Cuffiani. We thank A. Casoni for preparing this manuscript.

**REFERENCES**
[1]    G. Goldhaber et al, Phys. Rev. 120 (1960) 130.
[2]    Hanbury Brown et al, Phil. Mag. 54 (1954) 633; Nature 178 (1956) 1046.
[3]    M. Cuffiani and G. Giacomelli, Il Nuovo Saggiatore 18 (2002) 46, n. 1-2.
[4]    G. Abbiendi et al, hep-ex/0302027, Phys. Lett. B559 (2003) 131.
[5]    K. Ahmet et al, Nucl. Instr. and Meth. A305 (1991) 275.
[6]    M. Boutemeur, Nucl. Phys. B (Proc. Suppl.) 121/C (2003) 82. P.D. Acton et al., Phys. Lett. B267 (1991) 143. P.D. Acton et al, Phys. Lett. 298 (1993) 456. G. Abbiendi et al., Eur. Phys. J. C21 (2001) 23. R. Akers et al., Z. Phys. C67 (1995) 389. G. Abbiendi et al., Eur. Phys. J. C8 (1999) 559.
[7]    G. Alexander et al, Phys. Lett. B452 (1999) 159.
[8]    G. Alexander et al, Z. Phys. C72 (1996) 389.
[9]    G. Abbiendi et al, Eur. Phys. J. C16 (2000) 423.
[10]   K. Ackerstaff et al, Eur. Phys. J. C5 (1998) 239.